\documentclass{aastex631}

\usepackage{csquotes}
\usepackage{gensymb}
\usepackage{subfig}

\newcommand{\swift}{\textit{Swift}}
\newcommand{\foo}{\hspace{-2.3pt}$\bullet$ \hspace{5pt}}

\begin{document}

\title{Swiftly chasing gravitational waves across the sky in real-time}

%% The new \altaffiliation can be used to indicate some secondary information
%% such as fellowships. This command produces a non-numeric footnote that is
%% set away from the numeric \affiliation footnotes.  NOTE that if an
%% \altaffiliation command is used it must come BEFORE the \affiliation call,
%% right after the \author command, in order to place the footnotes in
%% the proper location.
%%
%% Use \email to set provide email addresses. Each \email will appear on its
%% own line so you can put multiple email address in one \email call. A new
%% \correspondingauthor command is available in V6.31 to identify the
%% corresponding author of the manuscript. It is the author's responsibility
%% to make sure this name is also in the author list.
%%

\author[0000-0002-2810-8764]{Aaron Tohuvavohu}
\affiliation{David A. Dunlap Department of Astronomy and Astrophysics, University of Toronto,
Toronto M5S 3H7, Canada}
\affiliation{Dunlap Institute for Astronomy and Astrophysics, University of Toronto, Toronto M5S 3H7, Canada}
\affiliation{Cahill Center for Astronomy and Astrophysics, California Institute of Technology,
Pasadena, CA 91125, USA}
\author[0000-0002-6745-4790]{Jamie A. Kennea}
\affiliation{Department of Astronomy and Astrophysics, The Pennsylvania State University, University Park, PA 16802, USA}
\author[0000-0002-8866-7891] {Christopher J. Roberts}
\affiliation{Astrophysics Projects Division, NASA Goddard Space Flight Center, Mail Code 440, Greenbelt, MD 20771, USA}
\author[0000-0001-5229-1995]{James DeLaunay}
\affiliation{Department of Astronomy and Astrophysics, The Pennsylvania State University, University Park, PA 16802, USA}
\author[0000-0003-0020-687X]{Samuele Ronchini}
\affiliation{Department of Astronomy and Astrophysics, The Pennsylvania State University, University Park, PA 16802, USA}
\affiliation{Institute for Gravitation \& the Cosmos, The Pennsylvania State University, University Park, PA 16802, USA}
\author[0000-0003-1673-970X]{S. Bradley Cenko}
\affiliation{Astrophysics Science Division, NASA Goddard Space Flight Center, Mail Code 661, Greenbelt, MD 20771, USA}
\affiliation{Joint Space-Science Institute, University of Maryland, College Park, MD 20742, USA}
\author[0000-0001-9178-5744]{Becca Ewing}
\affiliation{Department of Physics, The Pennsylvania State University, University Park, PA 16802, USA}
\author[0000-0001-9769-531X]{Ryan Magee}
\affiliation{LIGO Laboratory, California Institute of Technology, Pasadena, CA 91125, USA}
\author[0000-0002-8230-3309]{Cody Messick}
\affiliation{University of Wisconsin-Milwaukee, Milwaukee, WI 53201, USA}
\author[0000-0002-0525-2317]{Surabhi Sachdev}
\affiliation{School of Physics, Georgia Institute of Technology, Atlanta, GA 30332, USA}
\author[0000-0001-9898-5597]{Leo P. Singer}
\affiliation{Astrophysics Science Division, NASA Goddard Space Flight Center, Mail Code 661, Greenbelt, MD 20771, USA}

\begin{abstract}
We introduce a new capability of the Neil Gehrels \swift\ Observatory, dubbed `continuous commanding,' that achieves $10$ seconds latency response time on-orbit to unscheduled Target of Opportunity requests received on the ground. We show that this will allow \swift\ to respond to pre-merger  (early warning) gravitational-wave detections, rapidly slewing the Burst Alert Telescope (BAT) across the sky to place the GW origin in the BAT field of view at, or before, merger time. This will dramatically increase the GW/GRB co-detection rate, and enable \textit{prompt arcminute localization of a neutron star merger}. We simulate the full \swift\ response to a GW early warning alert, including input sky maps produced at different early warning times, a complete model of the \swift\ attitude control system, and a full accounting of the latency between the GW detectors and the spacecraft. 60 s of early warning can double the rate of a prompt GRB detection with arcminute localization, and 140 s \textit{guarantees} observation anywhere on the unocculted sky, even with localization areas $\gg 1000$ deg$^2$. While 140 s is beyond current gravitational wave detector sensitivities, 30-70 s is achievable today. We show that the detection yield is now limited by the latency of LIGO/Virgo cyber-infrastructure, and motivate focus on their reduction. Continuous commanding has been integrated as a general capability of \swift, significantly increasing its versatility in response to the growing demands of time-domain astrophysics. We demonstrate this potential on an externally triggered Fast Radio Burst, slewing 81 degrees across the sky, and collecting X-ray and UV photons from the source position $<150$ s after the trigger was received from the Canadian Hydrogen Intensity Mapping Experiment (CHIME), thereby setting the earliest and deepest such constraints on high energy activity from non-repeating FRBs. The \swift\ Team invites the community to consider and propose novel scientific applications of ultra-low latency UV, X-ray, and gamma-ray observations.

\end{abstract}

\section{Introduction} \label{sec:intro}
The merger of two compact objects provides a rich laboratory in which to explore the fundamental physics of space-time. If at least one of these objects is baryonic, the potential yield expands enormously, with accretion physics, ultra-relativistic jet launching, heavy-element nucleosynthesis, perturbed supranuclear matter, and more all emerging rapidly following the merger \citep{burnsreview}. Advanced gravitational-wave detectors have begun to detect these binary coalescences in their late inspiral and merger phases, providing merger times, source classification, and sky localizations \citep{gwtc3}. Mining the rich post-merger physics requires electromagnetic facilities to rapidly locate and study these phenomena, before the quickly evolving processes shut off, or fade below detectability. In one case this has been partially successful, with the GW detection of a binary neutron star merger (BNS, GW170817; \citet{170817gw,170817mma}), a gamma-ray burst (GRB) detected 1.7 s later \citep{170817fermi,170817integral}, the subsequent search and discovery of the associated kilonova at 11 hours post merger in the optical and then UV and IR (\citealt{170817kilonova}, and references therein), followed by detection of a synchrotron afterglow at 16 days after the merger in X-ray \citep{170817xray1,170817xray2,170817xray3} and radio \citep{170817radio1,170817radio2}. These observations, and follow-up over years since merger, have yielded a tremendous amount of information on the production of heavy r-process elements and the chemical enrichment of the Universe, the physics and structure of relativistic jets, and the equation of state of matter at supranuclear densities \citep{170817review}.

However, much was missed. Focused observations with humanity's most sensitive facilities did not begin until $\sim1$ day after the merger \citep{170817mma}. Despite tremendous efforts, significant uncertainty remains in the composition of the ejecta that drove the kilonova, and the mechanism driving the early emission, due to model degeneracy. Some of these degeneracies can only be broken with early observations, during the kilonova's rise within a few hours post merger \citep{arcaviearly}. Additionally, several hypothesized emission components emerge at significantly earlier times. These include fast thermal UV emission driven by the beta-decay of free neutron after $\sim880$ s \citep{shri2005,2015MNRAS.446.1115M,2021ApJ...921..161D}, UV and X-ray emission from a magnetar remnant driving fast winds and shock breakout \citep{combi,2022Univ....8..633W}, very early TeV emission from the forward shock in the jet \citep{2023A&A...678A.126B}, an optical flash on timescale of 10s of seconds from synchrotron emission driven by the reverse shock \citep{2023NatAs...7..843O}, and more. It is clear that there is compelling physics that presents itself only briefly, within the first s to hours post-merger. Current approaches to counterpart discovery and characterization mostly focus on the thermal (kilonova) emission (which rises on time-scales of hours), and involve time consuming wide-field optical searches and transient characterization \citep{ztfsearch,decam}. Even in the case of a well localized GW event, typical time to optical counterpart discovery and follow-up with more powerful telescopes is hours-to-days. This approach is not suitable for accessing the early time physics.

The ideal scenario would be to locate the BNS promptly, to a precision sufficient for immediate follow-up with our most powerful focused telescopes -- $\sim$ arcminutes field of view) (FOV) This would bypass the slow wide-field search stage, and allow access to the early time physics with sensitive observations across the electromagnetic spectrum. To do this, we should use the earliest known emission. The gamma-rays are produced within seconds of merger, and existing instruments can determine gamma-ray source positions to the required precision. However, the combination of instantaneous FOV, sensitivity and localization precision required is challenging to achieve. While all-sky gamma-ray instruments do exist (eg Fermi/GBM, INTEGRAL/SPI-ACS) their localization precision is comparable to, or worse than, the GW localizations. In contrast, arcminute-scale localizations for gamma-rays can be determined, but the various existent techniques to achieve this necessarily limit the field-of-view. Simultaneously, high sensitivity is required due to the large fraction of BNS mergers where the GRB jet is viewed off-axis, resulting in fainter gamma-ray emission.

The Burst Alert Telescope (BAT) \citep{bat} on the Neil Gehrels \swift\ Observatory \citep{swift} is the most sensitive GRB detector in current operation, and capable of regular 1-3 arcminute precision positions. \citet{guano} and \citet{nitrates} introduced new data products and analysis techniques for \swift/BAT that significantly ($\sim4$x) increase the detection rate, and sensitive volume, for faint GRBs. These capabilities have led to multiple searches, including real-time joint searches with the International Gravitational Wave Observatory Network (IGWN), with results providing the deepest upper limits to date on gamma-ray emission at the time of GWs \citep{o3gbmswift,o3swift,o4samuele}. However, \swift/BAT is a directional instrument, providing superior sensitivity and localization precision over `only' $\sim2.2$ steradians, or $\sim17\%$ of the sky. This limits the current detection yield at least as much as the sensitivity does, as binary neutron star mergers are known to be rare in the local universe. If only we knew where to point...

In parallel, the increasing sensitivity of the IGWN (particularly at low frequencies) has raised a unique opportunity, to detect and roughly localize binary neutron star mergers seen in GW data \textit{before merger} \citep{2012ApJ...748..136C}. These early warning searches work in the same manner as traditional matched filter searches on the GW data, but instead use templates truncated at several different final frequencies during the inspiral, equivalent to different times before merger. However, current sensitivities of the GW detectors allow confident detection no more than 70 s before merger, and with localization areas of $>\mathcal{O}(10^3)$ deg$^2$. Not many instruments are able to take advantage of such an opportunity. This would require sensitivity to the prompt emission produced in BNS mergers, response time of 10s of seconds across the sky, and an instantaneous FOV of $>1000$ deg$^2$. Here we investigate the capabilities of \swift/BAT for this purpose, and show that the spacecraft and instrument are in-principle capable of successfully utilizing an early warning detection to ensure counterpart discovery.

\swift\ was designed to solve the mystery of short gamma-ray bursts and their connection to neutron star mergers, to quote the eponymous song: 
\begin{displayquote}
    Swiftly swirling, gravity twirling\\
    Neutron stars \textbf{about to collide}. \\
    Off in a galaxy so far away, \\
    Catastrophic interplay, \\
    Roller coaster gamma-ray ride.\\
    Super bright explosion then,\\
    Never to be seen again,\\ 
    How are we supposed to know? \\
    How about a telescope rotation [ \swift\ is the satellite that swings]\\
    Swiftly onto the location [Onto those brightly bursting things]\\ 
    Of its panchromatic afterglow? [To grab the multi-wavelength answer of what makes them glow!]\\ 
    --Padi Boyd, Swift Song, 1999
\end{displayquote}
\swift\ has been slewing immediately and autonomously within seconds of emission time since launch, triggered by a GRB detected onboard. However its capacity for rapid response to external triggers has long been hampered by operational and command and control limitations. Here we complete the final prophesied evolution of \swift, swinging it rapidly across the sky to chase down neutron stars \textit{before} they merge.

In Section \ref{sec:timescales} we discuss the expected gamma-ray emission and the relevant time-scales necessary for observation. In Section \ref{sec:ewgwdata} we describe the detection of gravitational waves pre-merger, and discuss some challenges inherent to using these alerts. In Section \ref{sec:swiftscenarios} we formulate the problem of early warning response and describe some constraints of \swift\ that shape the solution space. In Section \ref{sec:trajectory} we describe a rapid trajectory optimization algorithm to quickly solve the optimal response to an early warning gravitational wave alert. In Section \ref{sec:results} we present results after applying this strategy to a large sample of early warning alerts. In Section \ref{sec:latency} we explore practical latencies that can reduce the amount of time \swift\ has to respond, and demonstrate the capabilities of a new \swift\ capability for $<10$ s TOO response. In Section \ref{sec:conclusions} we conclude with some suggestions for LIGO/Virgo/Kagra necessary to enable this science, and a broader invitation to the community to use the new capabilities developed for \swift.

\section{Emission Targets and Timescales}
\label{sec:timescales}
% What are we looking for and when do we have to look?

Binary neutron star mergers are known to produce prompt GRB emission shortly after merger. In the case of GW170817, this signal began $\sim1.7$ s after merger time with a duration of $\sim2.5$ s. Because the jet launch delay timescale itself is expected to be very short, the physical origin of the delay in the emission seen likely depends on propagation of the jet to the gamma-ray production site \citep{2020ApJ...895L..33B}. This will in turn vary with jet Lorentz factor and the circum-merger environment.  The prompt emergence of the signal, and its short duration, necessitate observation at, \textbf{not after}, merger time. However, there can also be longer-lived gamma-ray emission. 

Up to $\sim10\%$ of \swift-detected short GRBs (which are thought to result from compact object mergers; \citealt{paoloreview}) are found to have extended emission, lasting $\sim10-100$ s after the prompt phase \citep{2015MNRAS.452..824K}. The origin is unknown, but is often interpreted as continued energy injection from a short-lived (magnetar) post-merger remnant \citep{2014MNRAS.438..240G}. In this case, the shut-off slope of the extended emission can be connected to either the magnetic braking-driven spin-down or the gravitational wave-driven spin-down of the remnant \citep{2020ApJ...898L...6L}. The combination of GW and gamma-ray observations can thereby uniquely constrain the nature of the post-merger remnant. We note that \swift/BAT with its imaging capabilities is especially sensitive to weak, long, emission. This longer-lived emission raises the possibility to detect and localize the merger site even if BAT arrives on target up to $\sim10-100$ s after merger.

The recent observations of 170817-like kilonovae signatures embedded in the afterglows of long GRBs 211211A \citep{2022Natur.612..223R,2022Natur.612..228T} and 230307A \citep{2024Natur.626..737L,Yang:2023mqt} have allowed their BNS merger origin to be established (as opposed to the death of massive stars, thought to be the origin of most long-duration GRBs; \citealt{longgrbreview}). These bursts have durations of $\sim50$ and 200 s, respectively. Co-detection with GW and prompt localization and follow-up will be crucial to probe how to power such long emission given the energy budget available in BNS mergers. In this case, it may be possible to detect and localize the merger site even if BAT arrives at the target $\sim50-100$ s after the merger.

Additionally, there are a plethora of models that predict `precursor' emission before merger time \citep[e.g.][]{2001MNRAS.322..695H}. These involve some interaction of the two neutron stars before merger, either of their magnetospheres \citep{2016MNRAS.461.4435M,2022MNRAS.515.2710M}, or tidal strain \citep{2013ApJ...777..103T}, triggering various emission mechanisms. The total energy budget available to drive such processes increases rapidly as the neutron stars approach merger, so the emission is generally expected no more than 10 s before the merger time \citep{2024arXiv240216504L}. Detection of precursor emission provides one of the only ways to directly probe the material/nuclear properties of the neutron stars \citep{2021MNRAS.504.1129N}, or their magnetic fields \citep{2021ApJ...921...92B}, before the messy merger. The maximum achievable luminosity of these phenomena is substantially lower than the prompt GRB emission, challenging the capabilities of current gamma-ray detectors.\footnote{Radio would be a better bet \citep{2024arXiv240216504L}.} \swift/BAT is the most sensitive GRB detector currently operating, and would have the best chance at detection, assuming arrival on target $\sim10$ s before merger. 

\section{Early Warning Gravitational Wave Detections}
\label{sec:ewgwdata}
A BNS system produces gravitational wave emission throughout its lifetime, slowly shedding angular momentum, and inspiralling towards each other while increasing their orbital frequency. The earliest possible detection therefore depends on the low-frequency sensitivity of the detector. The time to merger $t_c$ as a function of gravitational wave frequency $f$ is
\begin{equation}
t_c = \frac{5}{256} \frac{c^5}{(G \pi f)^{8/3}} \cdot \frac{1}{\mathcal{M}^{5/3}}
\end{equation}
where $\mathcal{M}$ is the chirp-mass of the system \citep{time2merge}. The gravitational wave signal enters the operating band of the current ground based detectors at 10-30 Hz, depending on signal amplitude. 

\citet{demoEW} first demonstrated the practical capability to detect BNS signal candidates, and distributed alerts before merger, on a replay of Observing Run 3 (O3) data. Alongside the search itself, they also highlighted some cyber-infrastructure challenges associated with data transfer, calibration, and alert generation to meet the strict latency demands.  In O4 all four IGWN real-time compact binary coalescence (CBC) pipelines (gstLAL, pyCBC, MBTA, SPIIR) now operate early warning searches alongside their full bandwidth searches \citep{gstlalEW,2020ApJ...902L..29N, spiirEW, mbta}. These pipelines utilize a small number (4-6) of different discrete truncation frequencies (29-58 Hz) corresponding to different pre-merger times (60 - 9 s, assuming a 1.4-1.4 M$_\odot$ BNS system). These truncated templates are used in a matched-filter search, allowing signals to be detected in the late inspiral stage as they accumulate SNR in the detectors. Depending on the distance, sky position, and source masses of the BNS a confident detection (network SNR $>10$) can be achieved up to 70 s before merger in the case of a GW170817-like signal (40 Mpc), or up to 50 s before merger in the case of a GW190425-like signal (200 Mpc). 

\begin{figure}
    \centering
    \includegraphics[width=\linewidth]{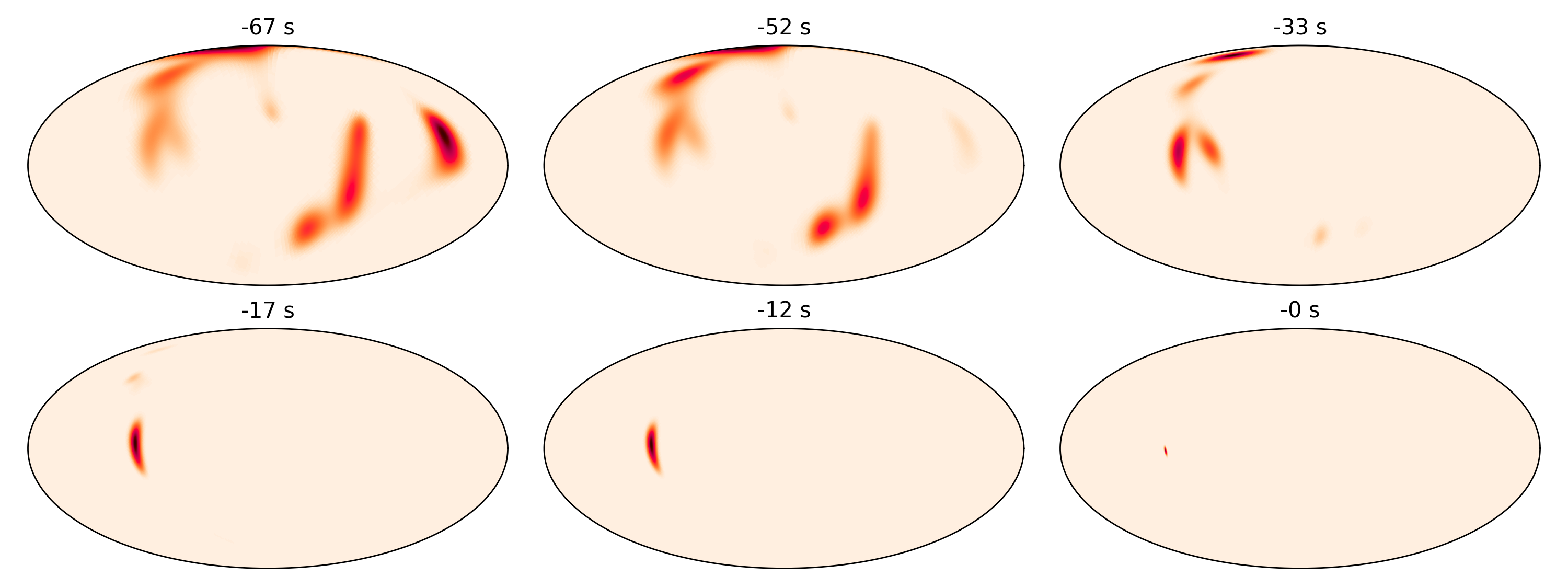}
    \caption[The time evolution of a GW sky map as a function of truncation frequency.]{GW localization sky maps at different truncation frequencies (pre-merger times) for a 1.4-1.2 M$_\odot$ merger at 50 Mpc. This example clearly shows the highest probability mode shifting between 3 locations in the sky, separated by $\sim180$\degree. In this case, a slew decision made on information from the earliest sky map could unknowingly result in looking at the wrong part of the sky.}
    \label{fig:GWevolution}
\end{figure}

Here we use the Early Warning Gravitational Wave Data Release \citep{gstlalEW} to study the characteristics of early warning detections and localizations, and the optimal strategy for early warning response with \swift. This data release comprises $\sim2$ million injected signals with known parameters representative of the expected BNS population demographics in the local Universe (to z = 0.2). Injections that are detected are tagged with their gstLAL-recovered SNRs, and BAYESTAR \citep{bayestar} produced 3D-sky maps at truncation frequencies of 29, 32, 38, 49, 56, and 1024 Hz corresponding to roughly 60, 45, 30, 15, 10 and 0 s before merger (for a fiducial 1.4-1.4 M$_\odot$ system). \citet{gstlalEW} present the localization area evolving as a function of truncation frequency, showing that the localization precision (characterized by 90\% enclosed area) evolves with time and accumulated SNR, slowly improving before rapidly converging in the last few s before merger. However, GW localizations can be multi-modal, with even relatively small sky areas split across opposite hemispheres. The maximum probability position can therefore shift back-and-forth across the whole sky before converging, as shown in Fig. \ref{fig:GWevolution}. 

\begin{figure}
    \centering
    \includegraphics[width=0.75\linewidth]{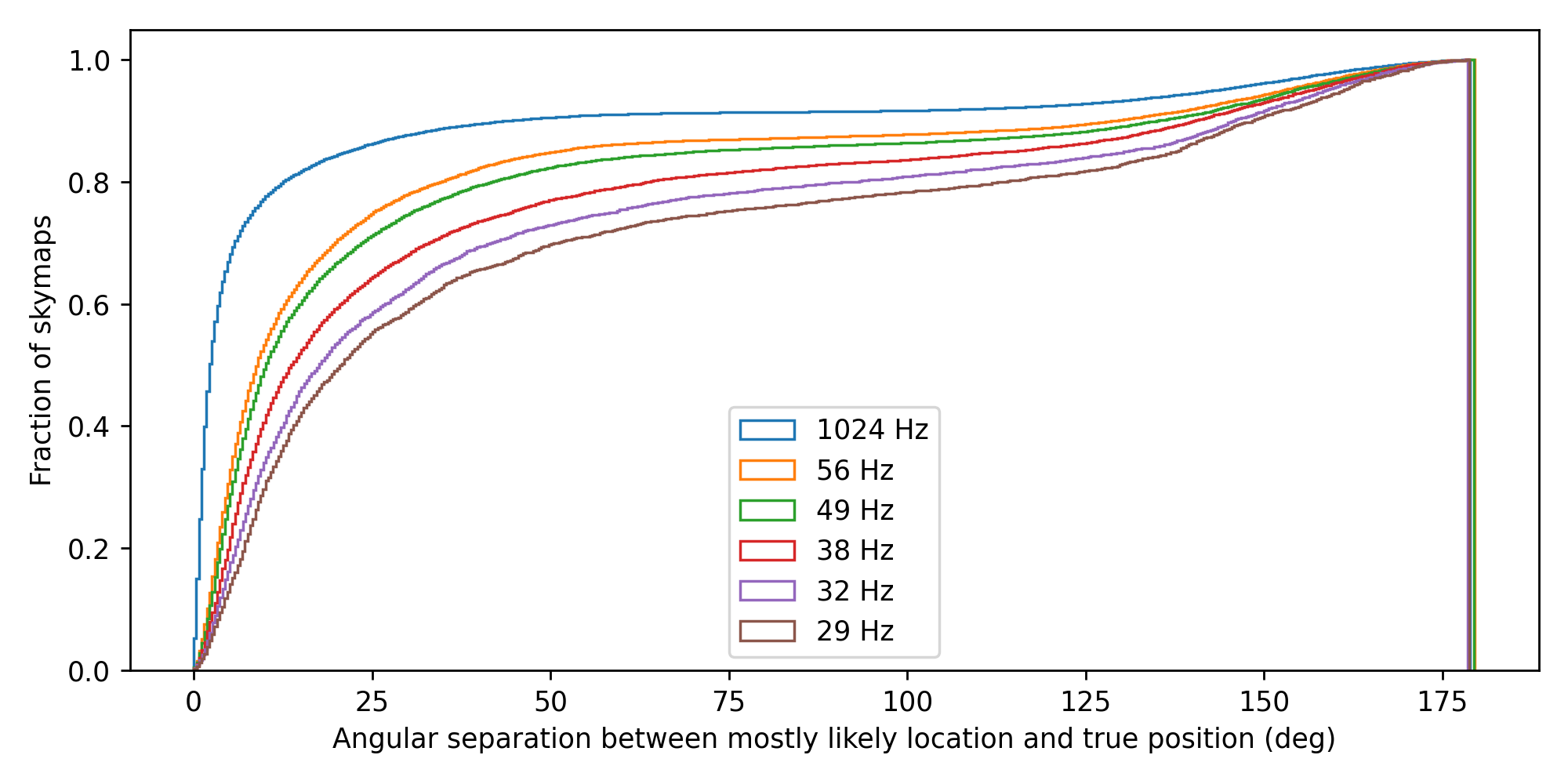}
    \caption[The position stability of the most probably mode on the GW sky map, as a function of truncation frequency.]{Cumulative distribution of the angular separation between the highest probability pixel in a sky map and the true location, for different truncation frequencies. An observer with a telescope of a given FOV radius can easily read off when to make a slew decision, depending on their risk posture and response time to target. The BAT FOV is oblong, but its semi-minor axis is 40\degree, thus guaranteeing coverage of the true position $>60\%$ of the time, \textit{if pointed successfully} at the highest probability location from the 29 Hz ($\sim-60$ s) sky map.}
    \label{fig:maxprobdist}
\end{figure}

Since this use case requires \textit{instantaneous} coverage at, or before, merger time by a telescope with finite FOV, a perhaps more relevant concern is the positional stability of the localization with time. In Fig. \ref{fig:maxprobdist} we compute the angular separation between the most probable sky pixel in each of the $\sim50,000$ early warning sky map, and the true (injected) position of the merger, as a function of truncation frequency.  We find that the true location is within 40\degree\ of the most probable sky map pixel $>60\%$ of the time at 29 Hz, and increasing for all higher truncation frequencies. In contrast, conditional on there being a BNS merger somewhere on the sky, it will land in the BAT FoV (semi-minor axis of 40 \degree) $\sim17\%$ of the time if we do nothing. To first order, this suggests that it is almost always beneficial to use the early warning information, modulo constraints. We will explore this more thoroughly in later sections.

We note that the Early Warning Data Release was produced assuming the LIGO and Virgo detectors at design sensitivity, which has not yet been achieved as of O4. This mostly affects the \textit{absolute} rate of expected detections and their distances, although changes to the low-frequency behavior of the detectors can and will shape the evolution of the sky localization with time. % Here we are interested here in the \textit{relative} yield improvement in GRB/GW co-detection that can be achieved by using the early warning alerts. 
Additionally, while these results pertain to detection prospects and parameter estimation achievable with realistic data, they do not include search pipeline or other cyber-infrastructure latency. We will return to this important issue in Sec. \ref{sec:latency}.

\section{Swift Observing Scenarios}
\label{sec:swiftscenarios}
We are interested in the optimal \swift\ response to the following question. \textbf{ If a BNS early warning alert is received at T0-X s until merger, with sky map Y, and \swift\ is currently observing target Z, what is the action that maximizes the chance of the true BNS location falling inside the BAT FOV at merger time?}

The response to this question can take the following forms: do nothing (high confidence that the GW origin is in the current BAT FoV), wait for improved information (expect more information in low latency that can substantially inform response), or immediately slew along a trajectory that will maximize the BAT/GW coverage at T0. Some specific characteristics and constraints of the \swift\ spacecraft and BAT shape the optimal approach:

\textbf{Once \swift\ begins to slew, it cannot be interrupted and redirected to another location until the slew is complete.} This has significant consequence in decision making under uncertainty, with the added expectation that updated information may (or may not) arrive. Given the average duration of \swift\ slews until settling, this lockout period can be $>100$ s. Since the GW sky maps evolve with time, and can contain multiple modes spanning a large fraction of the sky, a plausible scenario is that one is received, a slew is initiated towards the most probable area, and then an updated sky map arrives showing a refined localization on the other side of the sky. Ideally, we would like to avoid that.

\textbf{The BAT cannot trigger while slewing, but can collect data for analysis on the ground.} The BAT onboard triggers cannot run while the spacecraft is slewing, and are only activated once the `settled' flag is received from the attitude control system (ACS). This means that a typical on-board trigger at $\sim \mathrm{T_{GRB}}+6$ s, and associated alerts, are not feasible during these times. However, data can be collected during slews and promptly sent to the ground for rapid analysis \citep{guano}, and many GRBs are regularly detected during slews in this way \citep{slewgrb1,slewgrb2,slewgrb3,slewgrb4}. Indeed, over longer exposure times analyses performed on data taken during slews are actually more sensitive than inertially-pointed data, as the slew breaks degeneracies in the projected mask pattern, thereby reducing systematic noise in the reconstructed sky images \citep{copete}.

\textbf{The BAT coded FOV has significant rotational asymmetry.} The coded FOV is the sky area inside which BAT can provide arcminute localizations for detected bursts. BAT is capable of detecting bursts outside of the coded FOV with a reduced sensitivity, but is unable to provide localizations to arcminute precision for these bursts. This coded FOV is $\sim$ 120 x 80 degrees at its maximal dimensions (but it is not rectangular, see eg Fig. \ref{fig:pathology}) thereby implying a strong roll-angle dependence on the sky coverage for a given boresight pointing. Only a subset of roll-angles are available for a given target boresight at a given time, constrained by spacecraft capabilities. Since GW localization regions are also highly asymmetric, this can significantly affect feasible coverage.

\textbf{We do not have direct control over the precise \swift\ slew trajectory.} The autonomy onboard \swift\ and its attitude control system allows significant operational flexibility and substantially reduces risk. However, it also constrains some capability. When given target coordinates, \swift\ determines its own slew path that the ACS deems optimal and safe, which will depend on the instantaneous orientation of observing constraints (such as the Sun, Moon, Earth limb, and others). Some theoretically possible slew trajectories are therefore unavailable for use, and any specific trajectory desired needs to be reverse engineered from the target coordinates that would generate it at a given time from a fixed origin. Additionally, the velocity profile (of critical importance!) depends sensitively on the target position and roll angle.

\label{sec:trajectory}
\begin{figure}[!ht]
    \centering
    \subfloat{\includegraphics[width=0.5\textwidth]{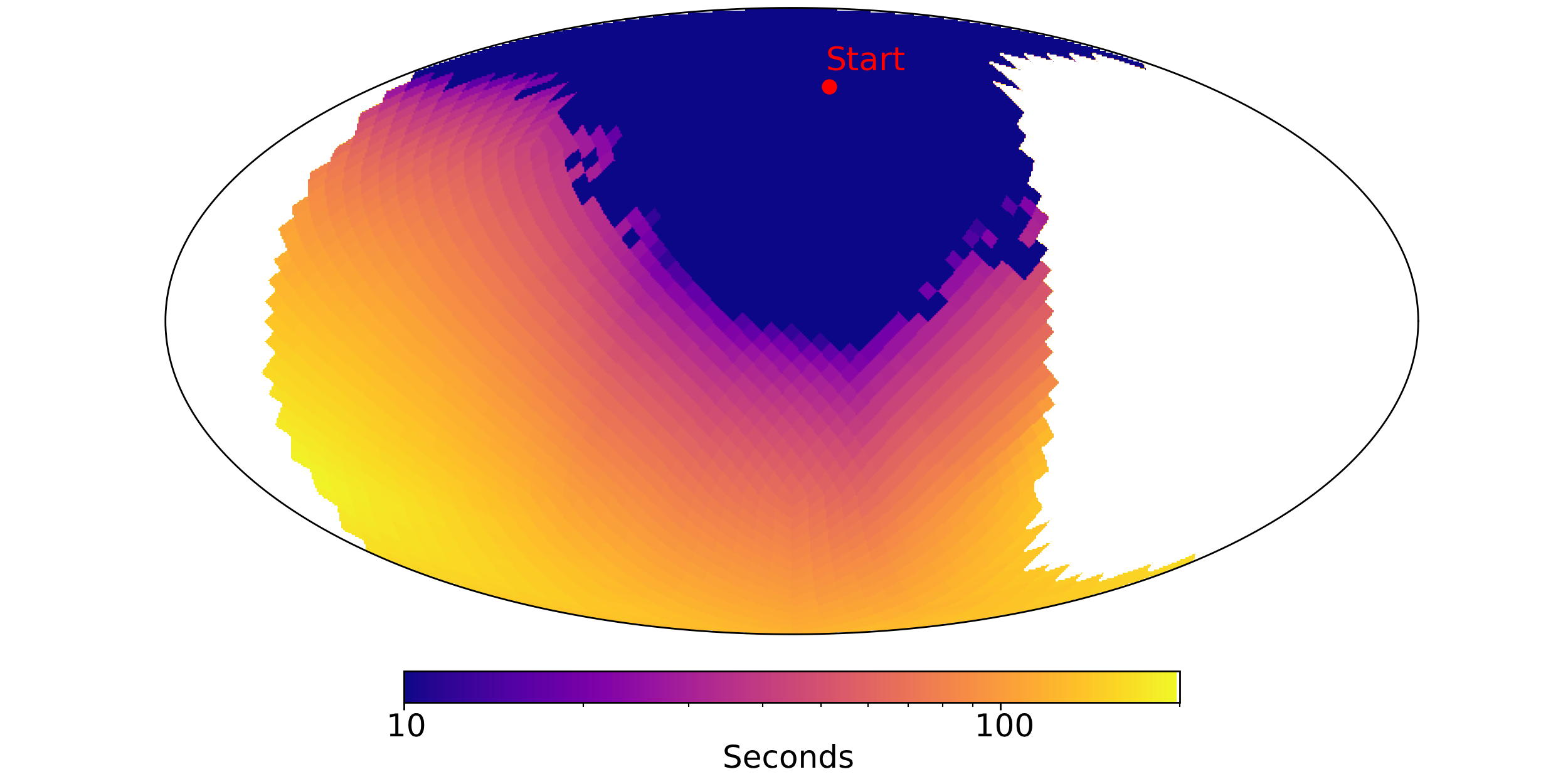}} 
    \subfloat{\includegraphics[width=0.5\textwidth]{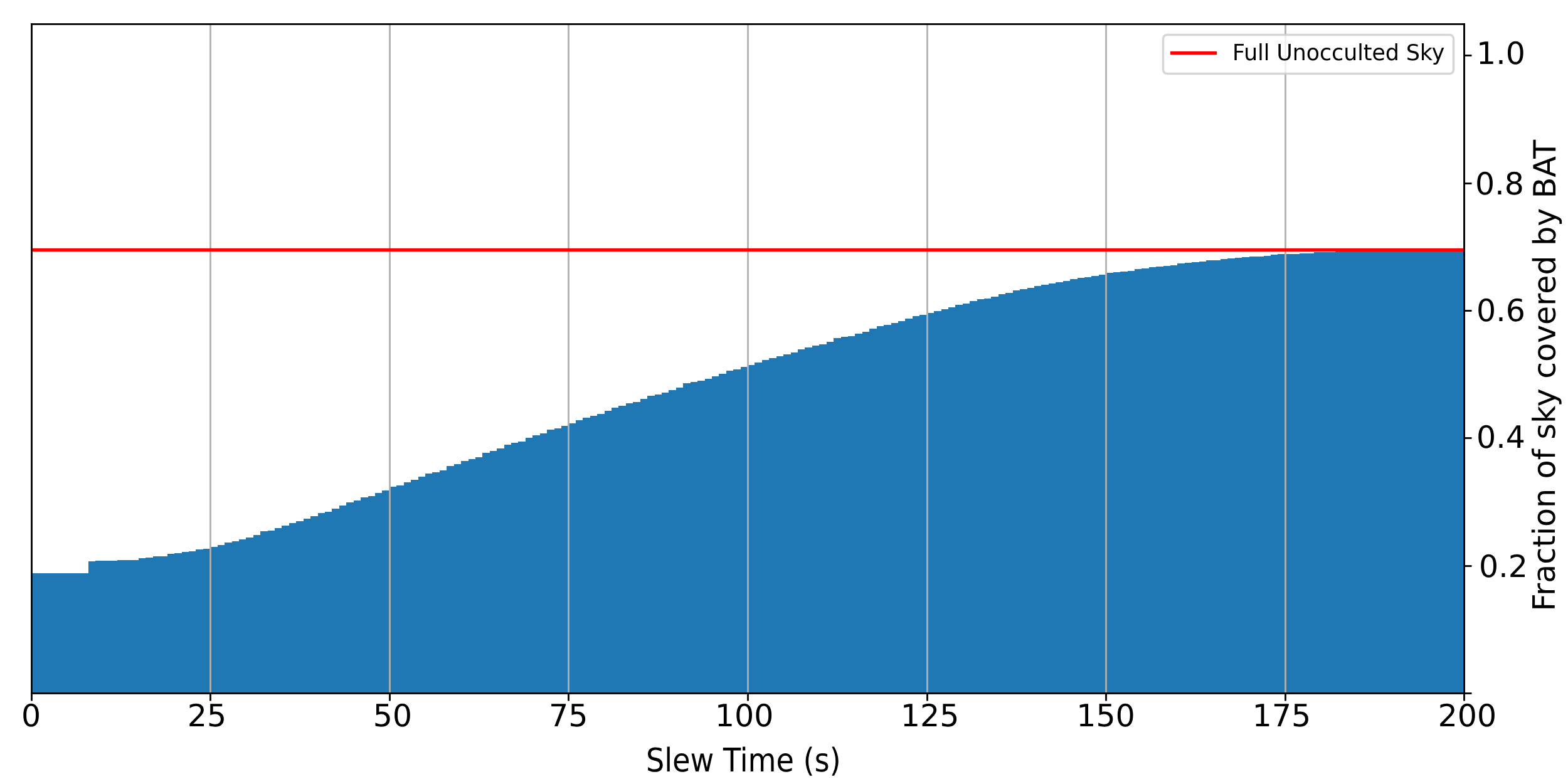}}
    \caption[Slew time to put any position on the sky into the BAT FOV.]{\textit{Left:} Minimum slew time required to put any location on the sky inside the BAT FOV, white sky areas are occulted by the Earth. \textit{Right:} The cumulative fraction of sky accessible to the BAT FOV as a function of slew time. If given a target, \swift\ can put almost any position on the unocculted sky into the BAT FOV in $<175$ s.}
    \label{fig:slewtimes}
\end{figure}

\section{Making the Right Choice: Trajectory Optimization}
From the above considerations, it is clear that the trajectory \swift\ takes as it slews to a new position will strongly impact the rate of prompt BNS detections. Here we examine the optimum slew trajectory that maximizes the fraction of the GW localization posterior contained inside the BAT coded FOV at merger time.
The \swift\ ACS was designed with significant redundancy, with 6 reaction wheels arranged for mutually orthogonal 3-axis control providing torque for rapid slews across the sky. In January 2022 one of these reaction wheels failed, and \swift\ now operates in a 5-wheel mode reducing the maximum torque available. In this mode, the maximum slew rate is $\sim0.8$ deg/s, but this maximum rate and maximum acceleration are not always achievable and depend strongly on the orientation of constraints on the sky and the roll angle change during the slew. Additionally, the \swift\ boresight must avoid transgressing within 27 degrees of the Earth limb, 45 degrees from the Sun, and 22 degrees from the Moon for instrument safety during slews. As such, the minimum slew distance between two points can often be substantially longer than the great circle distance. We note that the Earth limb moves $\sim3.8\degree$ per minute from the perspective of \swift, but slew safety constraints are evaluated by the ACS at the beginning of slews, not the end. Improper trajectory evaluation can therefore lead to \swift\ getting trapped between two constraints. These, and other similar vagaries, necessitate careful modeling when attempting to optimize the trajectory. In what follows, we use a complete model of the \swift\ ACS in its 5-wheel mode to achieve this.

In the general case it is not sufficient to simply solve for the \swift\ attitude at merger time that maximizes the GW sky map containment in the BAT FOV and then slew towards this. In most cases, the strictly optimal attitude is not achievable by merger time, and the attitude as a function of time along that trajectory will often not have optimal coverage at merger time. That is, \textit{the path to the optimal attitude at merger is not always the optimal trajectory}. This, combined with the fact that we can not directly shape the \swift\ trajectory, means that fully exploring the space of allowable \swift\ trajectories is required. In Fig. \ref{fig:slewtimes} we show an example of the minimum time required to put any location on the unocculted sky into the BAT FOV by exhaustively searching over all possible trajectories, a computationally expensive task ill-suited for real-time operations.

\begin{figure}
    \hspace{-1.65cm}
    \includegraphics[width=1.18\linewidth]{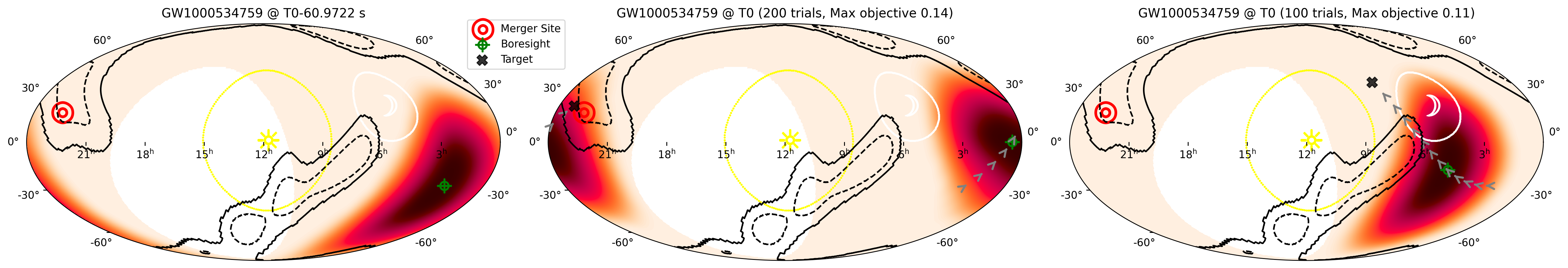}
    \caption[Optimal slew trajectory for a given GW early warning alert vs the trajectory found in 100 trials.]{A GW early warning alert at 29 Hz. Plotted are the GW sky map 50\% and 90\% containment contours and BAT FOV (heatmap) at the exact truncation frequency time, relative to merger T0. Left shows the orientation at time the alert is produced, with little coverage of the GW sky map. Center and right show the optimal trajectory found by the sampler (after 200 and 100 trials respectively), and the BAT FOV at merger time if that trajectory is executed. The portion of the sky occulted by the Earth at the time is in white, and the Sun and Moon boresight constraints are marked. The bulls-eye marks the true location of the merger, not known to the trajectory optimizer. The black X is the commanded target of the slew, and the green reticle is the \swift\ boresight.}
    \label{fig:pathology}
\end{figure}

The response to early warning gravitational wave alerts sets strong requirements on latency. We have allocated \textit{1 s of latency between alert receipt and target uplink to the spacecraft}. This includes processing the alert, determine the response strategy, and finding the optimal trajectory and the target that will achieve it. Meeting this latency requirement necessitates determining the optimum trajectory very rapidly. We formulate an objective function \texttt{slewmaxcorr} which takes as hyperparameters the slew target RA, Dec, Roll, and fixed parameters of t$_{\mathrm{start}}$, time until merger, and the GW sky map. It then computes the full time-resolved slew trajectory beginning at t$_{\mathrm{start}}$, and finds the instantaneous \swift\ attitude along this path at merger time. The inner product between the BAT FOV coverage at this time and the GW sky map (masked for occultation) is then calculated, and returned. Hereafter we refer to this quantity as the objective value and its maximization identifies the optimal Swift trajectory. We maximize \texttt{slewmaxcorr}, sampling over the space of allowed target coordinates (constrained samples are pruned). \texttt{slewmaxcorr} is optimized for speed, but still requires $\sim10$ milliseconds to evaluate per trial.  This implies we need to find an optimized solution in $\leq 100$ trials in order to meet our total latency requirements.

\begin{figure}
    \centering
    \includegraphics[width=0.8\linewidth]{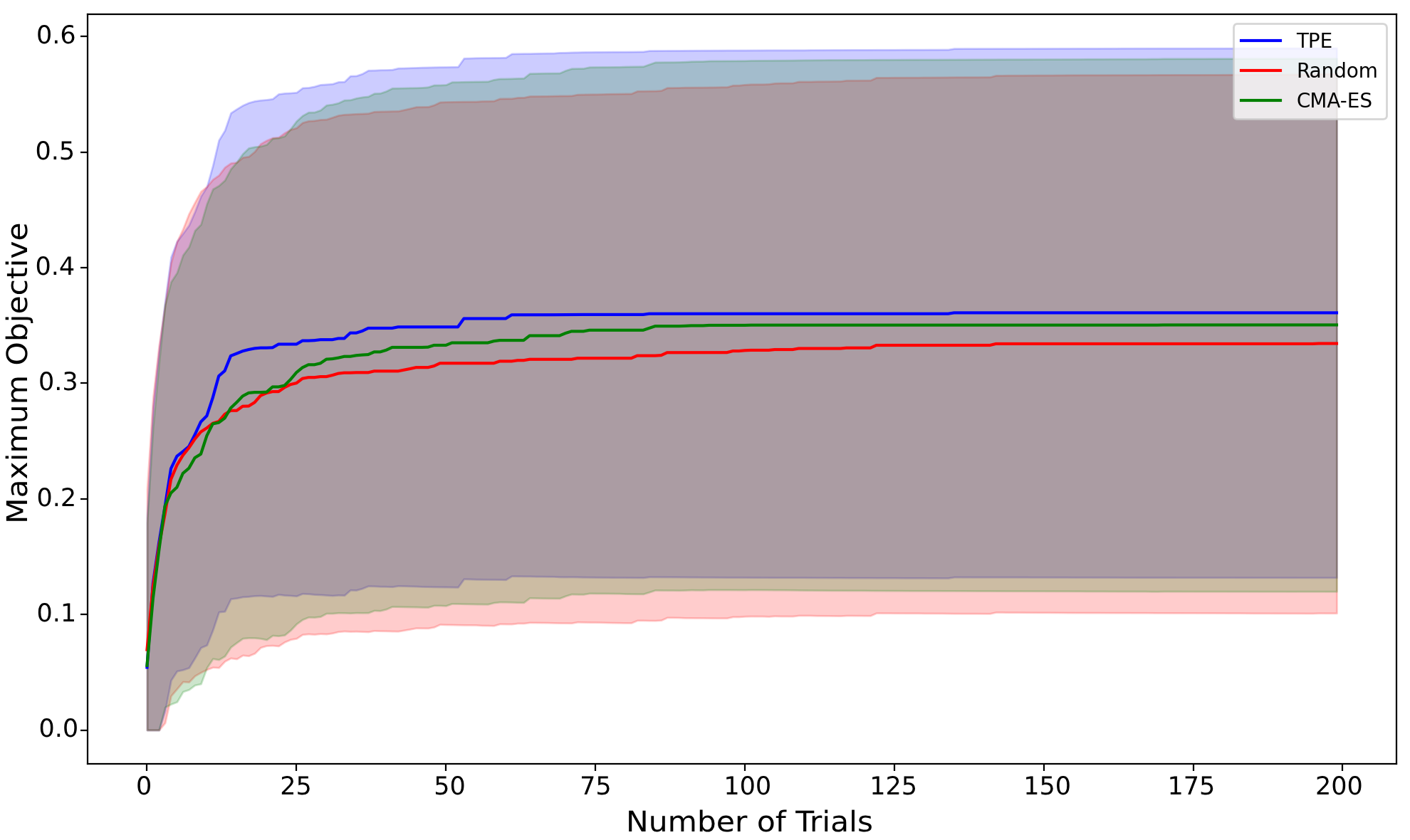}
    \caption[Comparison of the performance of 3 different samplers for exploring the space of optimal trajectories]{Comparison of  best objective values achieved by the TPE, Random, and CMA-ES samplers as a function of trials. The shaded areas represent the standard deviation across $>1000$ runs on different sky maps and starting conditions. TPE shows slightly superior overall performance and consistency, achieving the best results on average, although all samplers show significant performance variability across trajectory conditions.}
    \label{fig:compare-samplers}
\end{figure}

The combination of operational constraints, non-linear slew dynamics, irregular GW sky maps, and non-uniform sensitivity across the FOV leads to a highly non-convex and multi-modal optimization problem where traditional gradient-based methods can easily get trapped in local optima. We employ a global optimization strategy with probabilistic sampling and trial pruning. We utilize the Tree-structured Parzen Estimator \citep[TPE][]{TPE}, which works by constructing probabilistic models to efficiently explore the search space, allowing it to focus on areas with higher potential for improvement while still exploring broadly. This probabilistic approach enables TPE to effectively guide the optimization even in complex, non-convex spaces. We see this performance in Fig \ref{fig:compare-samplers}, where we compare TPE to a random (independent) sampler, and a Covariance Matrix Adaptation Evolution Strategy (CMA-ES) sampler on a wide range of input conditions on trajectory optimization. On average TPE reaches better objective values after 100 trials, although all samplers show substantial variance. This makes TPE the best choice when considering both the speed of convergence and the quality of the solutions found within a fixed number of trials. Convergence is typically achieved within 100-150 trials, although pathological cases do exist. For the specific case depicted in Fig. \ref{fig:pathology} a slow convergence means that the merger is not successfully observed, when triggering a slew after 100 trials. For this case the commanded final roll angle has greater relative `importance' compared to the RA, Dec, due to its strong impact on the maximum slew velocity, and the starting position relative to the multi-modal sky map. We note that additional sampler tuning, transfer learning from sub-samples of the simulation set, and further optimization of the objective function are likely to improve performance, particularly for these edge cases. 

In the following section we utilize the TPE trajectory optimizer, which is allowed to maximize the number of trials performed under a timeout of 1 second, to match the performance requirements of the real-time early warning response system.

\section{Simulations of \swift\ Response to GW Early Warning Alerts and Projected Yields}
\label{sec:results}

\begin{figure}
    \newpage
    \centering
    \includegraphics[width=\linewidth]{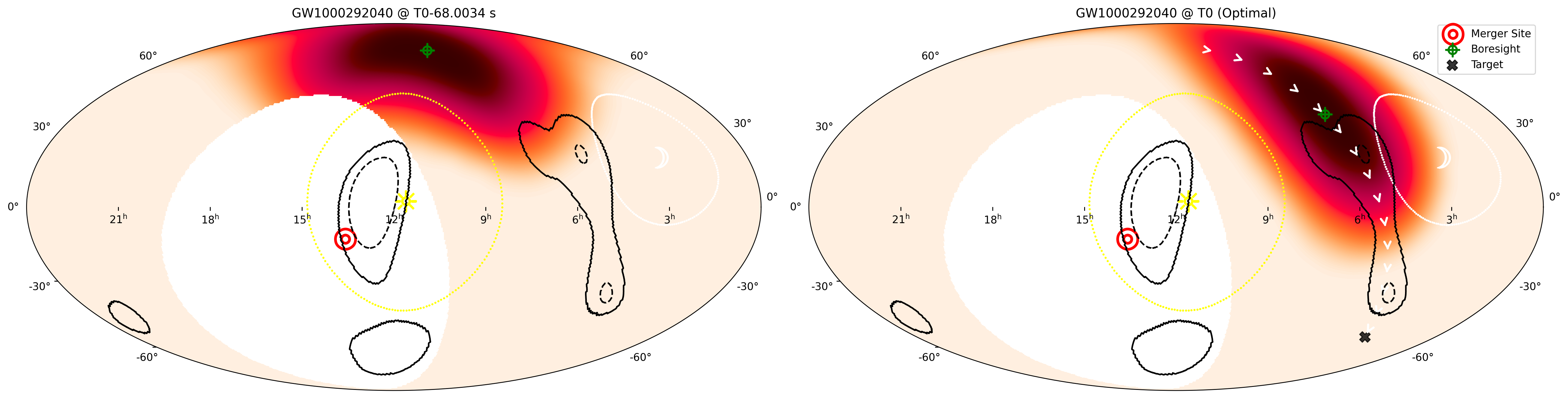}\\
    \includegraphics[width=\linewidth]{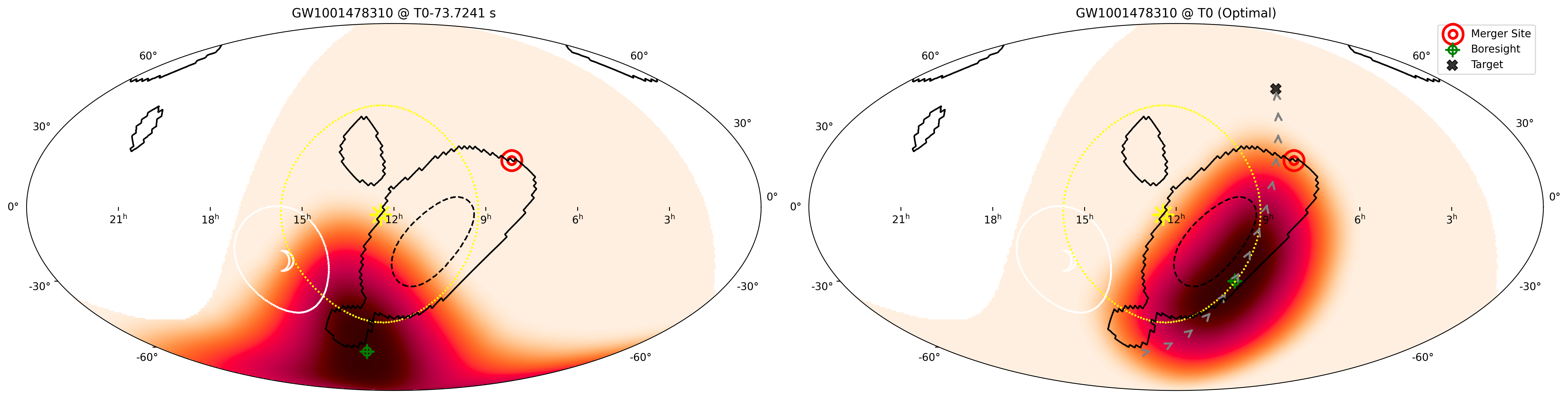}\\
    \includegraphics[width=\linewidth]{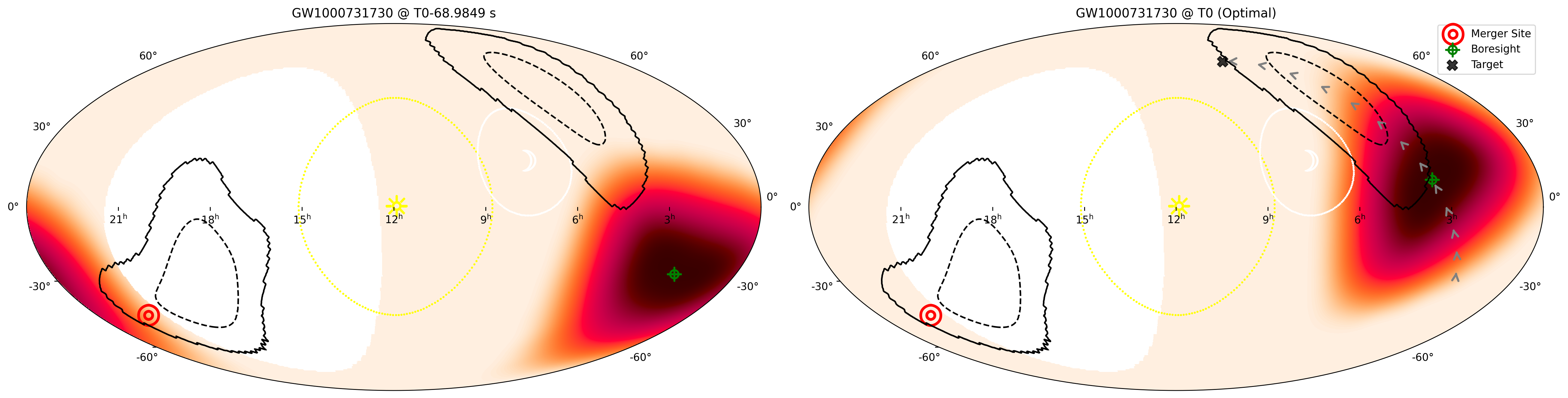}\\
    \includegraphics[width=\linewidth]{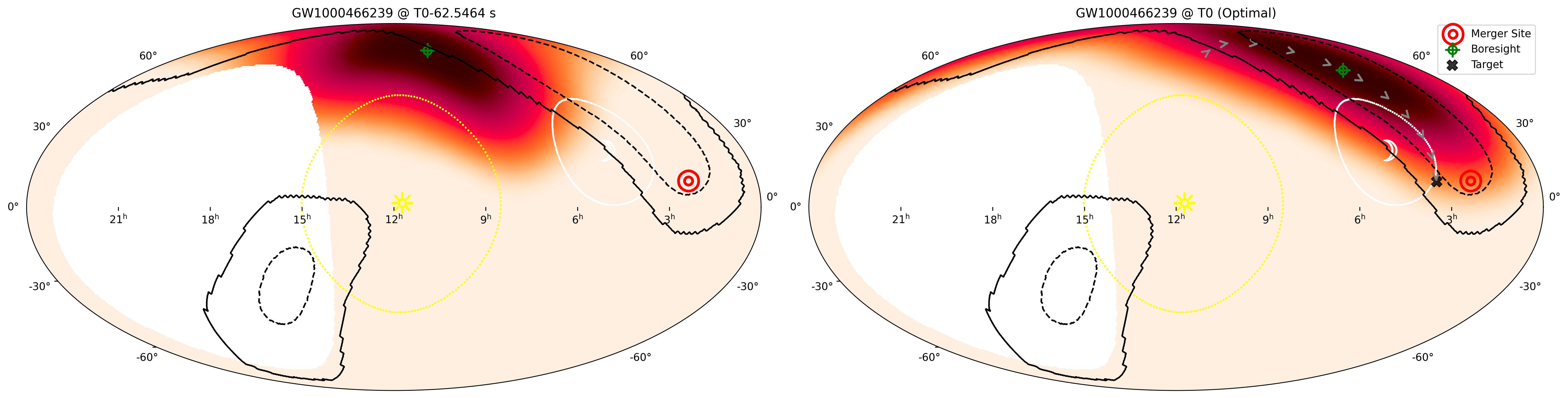}
    \caption[Sample of early warning GW response trajectories for truncation frequency of 29 Hz.]{A sample of GW early warning sky maps at truncation frequency of 29 Hz ($\sim-60$s before merger). Left column shows BAT coverage at alert time, right panel shows optimal trajectory found and BAT coverage at merger time if the trajectory is executed. Note a tragic (rare) occurrence in the 3rd row of slewing away from the correct position, not yet known. But note also marked successes in the 2nd and 4th panels.\\
    \\
    }
    \label{fig:29hz-response}
\end{figure}

\begin{figure}
    \centering
    \includegraphics[width=\linewidth]{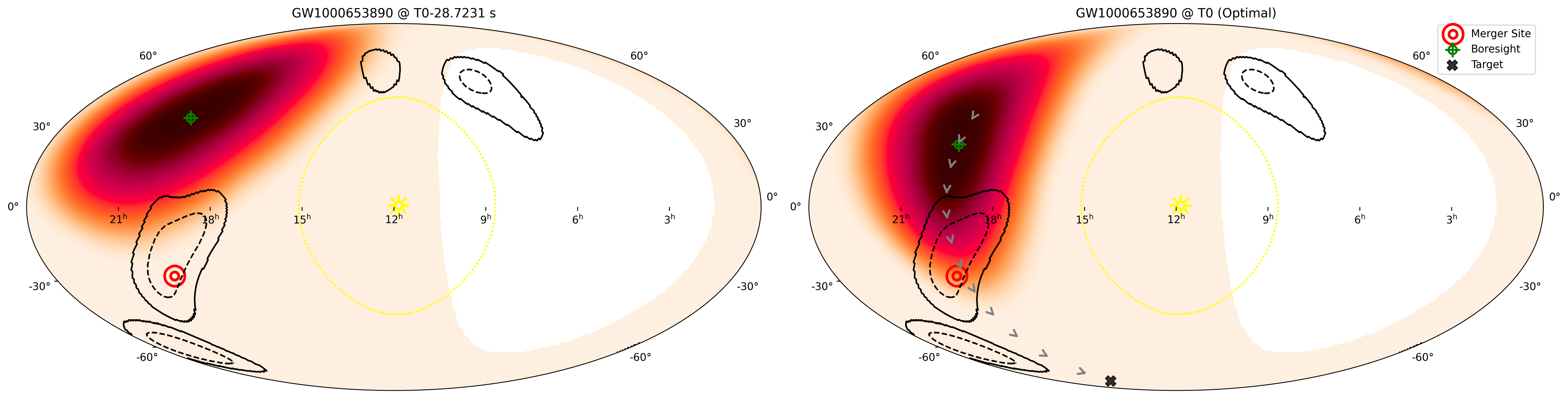}\\
    \includegraphics[width=\linewidth]{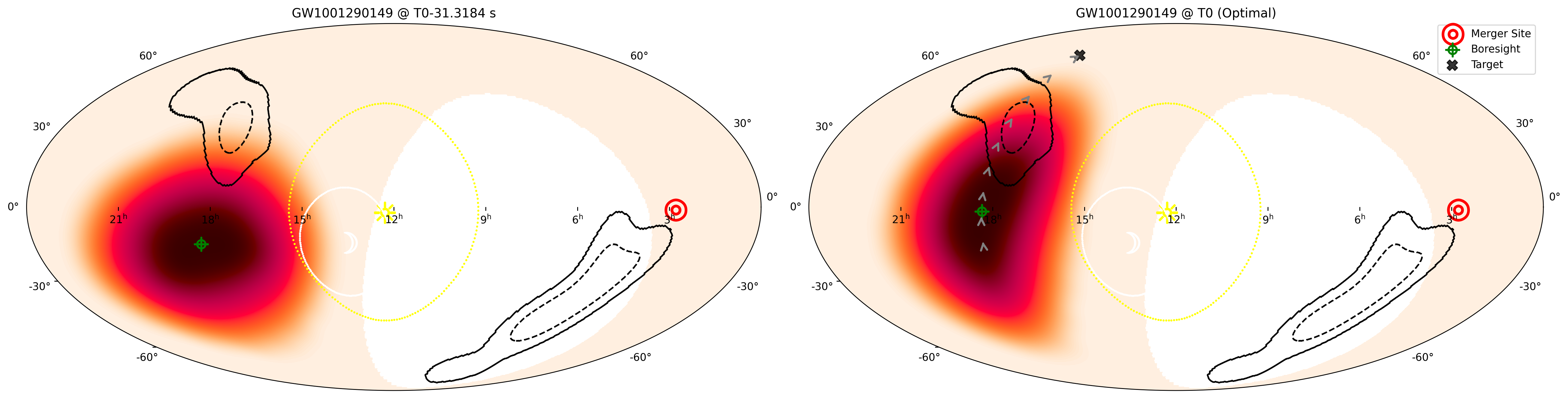}\\
    \includegraphics[width=\linewidth]{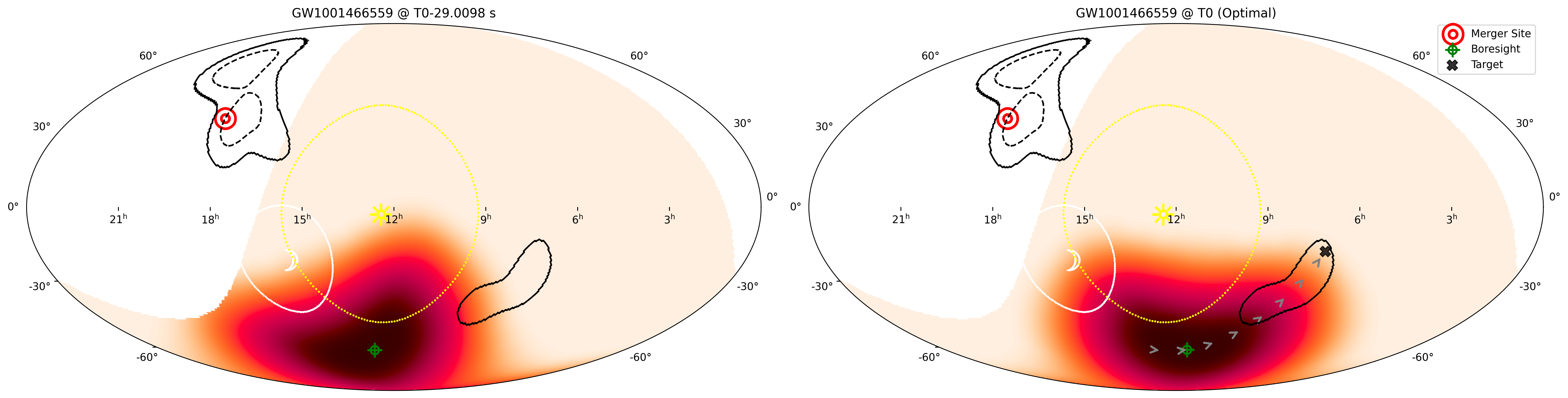}\\
    \includegraphics[width=\linewidth]{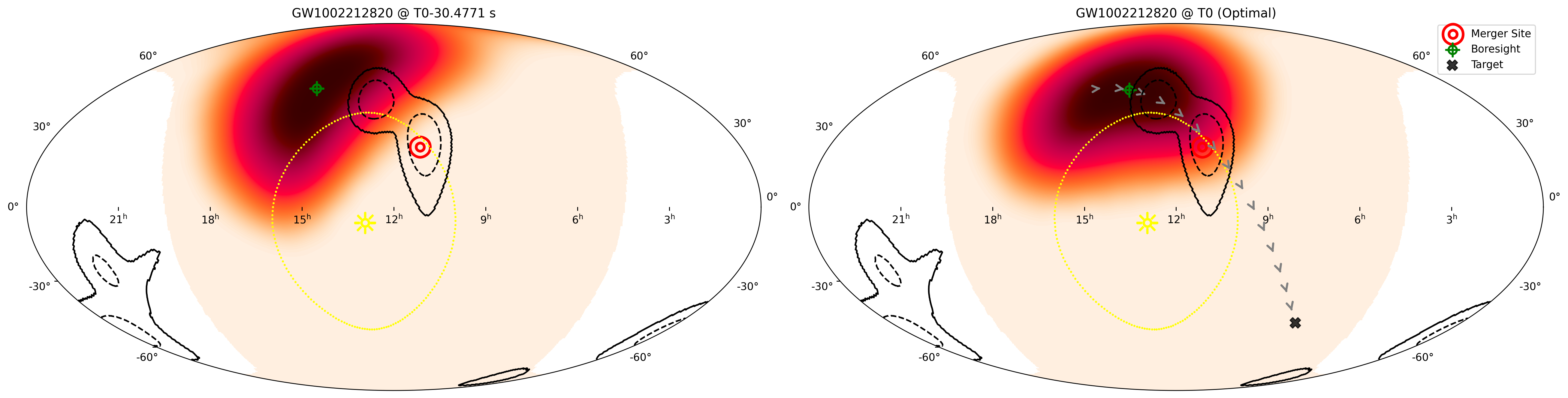}
    \caption[Sample of early warning GW response trajectories for truncation frequency of 38 Hz.]{Same as Fig. \ref{fig:29hz-response}, but for truncation frequency of 38 Hz. Note the successes in the first and fourth panel.}
    \label{fig:38hz-resposne}
\end{figure}

With a response strategy in hand, we aim to faithfully reproduce the actions that \swift\ will take upon receiving an early warning alert, and study the expected yield. From the early warning data release, 2737 (4131) [7775] injections have detections at 29 (32) [38] Hz. For each detection we create an instantaneous GW alert and ingest it as we would in real-time. For each early warning alert we project the sky maps onto the sky as seen by \swift\ at the time of the alert. In order to preserve the relationship between time, sky location, SNR, and localization precision we retain the true time of the signal, and calculate the actual orbital track, time-resolved orientation and size of observing constraints and occultations, and the as-flown observing schedule of \swift. We compute the effective sensitivity of BAT to the true position of the BNS (known from the injection), and find that $\sim17\%$ of all mergers will serendipitously occur inside the BAT FOV, consistent with expectations from the size of the FOV (2.2 steradian).\footnote{Note that this implies that the \swift\ observing schedule is indeed completely uncorrelated with the IGWN antenna pattern.} We also find that $32\%$ will be occulted by the Earth at merger time, consistent with the fraction of sky blocked by Earth at \swift's orbital altitude. We then apply the trajectory optimizer described in Sec. \ref{sec:trajectory}, and obtain a target RA, Dec, Roll to send to the spacecraft. We propagate the slew resulting from this target command, and find the position and orientation of the BAT FOV on the sky at merger time for a range of slew start latencies from 0 to 60 s. As such this latency fully encapsulates all in-principle reducible latencies, including IGWN pipeline latencies and \swift\ command latencies, as will be discussed further in Sec. \ref{sec:latency}, but assumes fixed detector sensitivity and \swift\ ACS. We again compute the effective sensitivity of BAT to the true position of the BNS at merger time, and record these result. See Figs \ref{fig:29hz-response} and \ref{fig:38hz-resposne} for representative samples from this simulation set.

\begin{figure}
    \hspace*{-1.5cm}
    \subfloat{\includegraphics[height=6cm,width=0.7\linewidth]{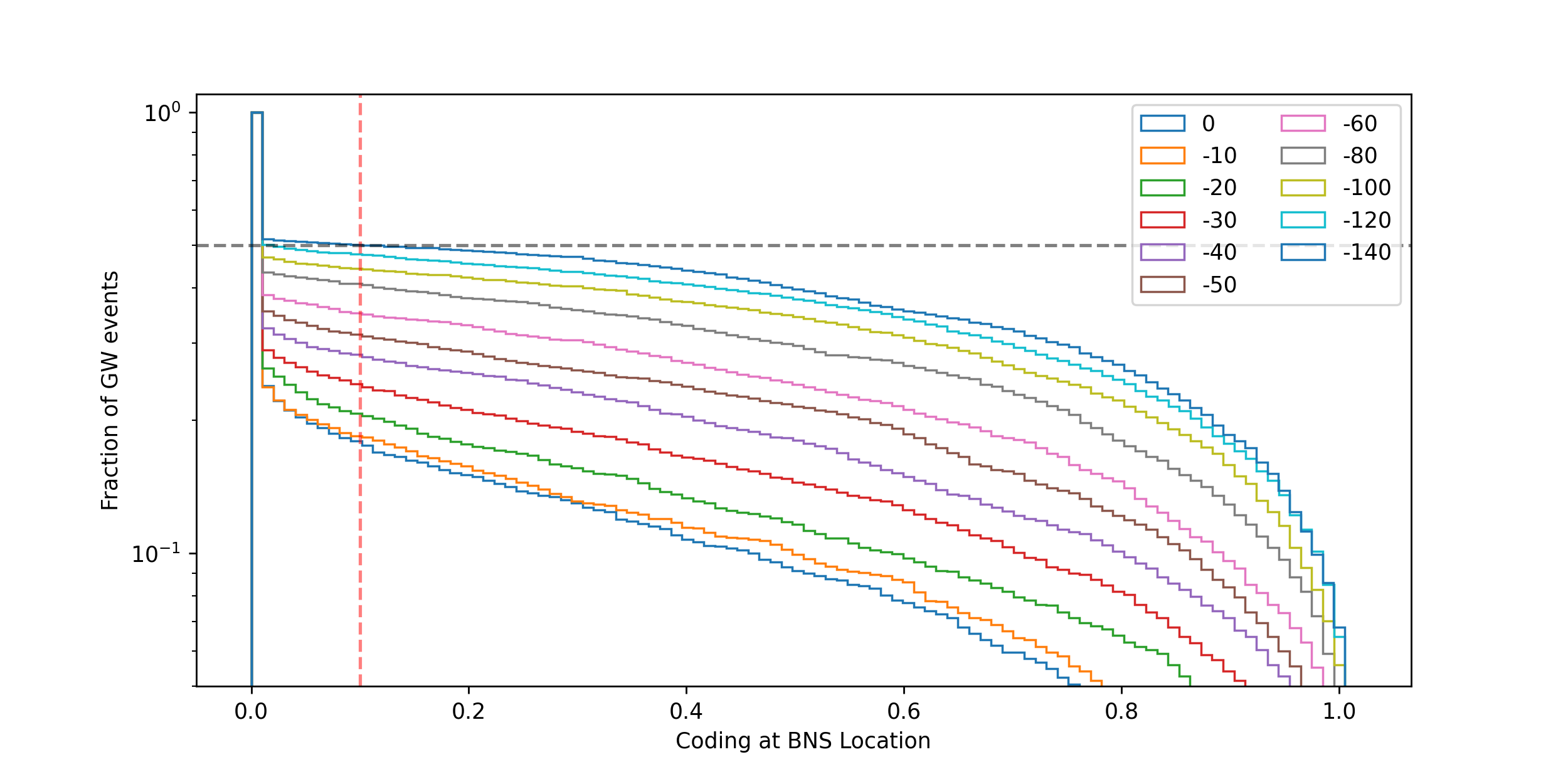}}
    \hspace*{-1.5cm}
    \subfloat{\includegraphics[height=6cm,width=0.5\linewidth]{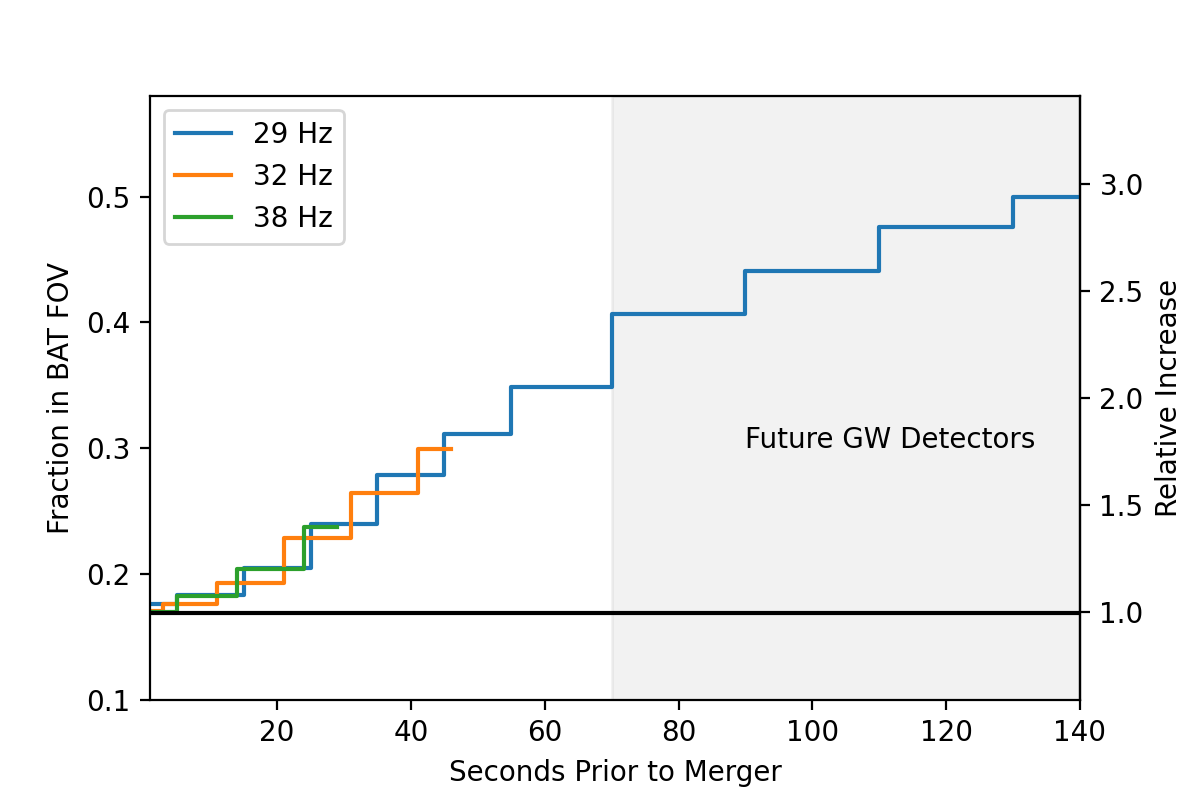}}
    \caption[Distribution of BAT sensitivity to the true neutron star merger position after executing an early warning response, as a function of latency.]{The reverse cumulative distribution of the BAT normalized sensitivity to the true BNS location, using 29 Hz sky maps at various response latencies. We mark the saturation condition for 29 Hz at 50\% of mergers, and the `edge' of the BAT FOV at 0.1. One can easily read off that early warning time of -60 s (pink) doubles the fraction of mergers inside the FOV. Note that the relative gain increases with coding, meaning that early warning response not only maximizes the fraction of mergers in the FOV, but also independently increases the average sensitivity to those already in the FOV. On the right, we present the fraction of mergers inside the BAT FOV, as a function of warning time for sky maps derived at different truncation frequencies. We can easily see that, despite the improved localization precision at higher frequencies, acting on the earliest sky map is always better than waiting given \swift's finite slew speed. We also show capabilities assuming early warning time $>70$ s, achievable only by upgraded GW detectors. For these cases we assume the same sky maps as produced at 29 Hz by Avanced LIGO/Virgo.}
    \label{fig:results}
\end{figure}

We define our success as the true location of the BNS landing inside the BAT coded FOV, where arcminute precision positions can be found for detected GRBs. The BAT coded FOV does not have uniform sensitivity, and we define the edge as the 10\% coding fraction, although brighter bursts are regularly localized at lower coding. BAT imaging effective area is $\sim$ linear in coding fraction. The results for response with 29 Hz sky maps are shown in Fig. \ref{fig:results} left, where the reverse cumulative fraction of BAT coding fraction at the true BNS location is shown, as a function of time between slew start and merger.

Using sky maps produced at 29 Hz, we expect early warning response to saturate our success rate at $\sim50\%$ of all mergers (a $3\times$ relative improvement), even at 0 latency and with an optimal trajectory. This is because $30\%$ of the time the true position will be blocked by the Earth, and an uncorrelated $\sim35\%$ (30\%) of the time we will look in the wrong direction, given the localization precision at 29 (32) Hz and the size of our field of view (see Fig. \ref{fig:maxprobdist}). We note that the Earth cannot be made smaller, nor \swift\ raised in its orbit, so the only way to raise this success ceiling is to improve the GW detector sensitivity such that sky localizations are better constrained at low truncation frequency.  When utilizing higher truncation frequency (less warning) sky maps, the precision improves, but the trade is for less warning time. We show that using the earliest sky map available is the correct choice given the \swift\ slew speed, in the right panel of Fig. \ref{fig:results}.

Our results show that \textit{with zero latency between the GW signal arriving at 29 Hz, and \swift\ beginning to slew, we can more than double the rate of mergers falling inside the BAT FOV, and of prompt arcminute localizations.} But, there's no such thing as a zero latency system. In Fig. \ref{fig:results} we see the strong impact of latency on the success rate, and in the following section we investigate the real-world latency of the various infrastructure between the wiggling LIGO mirrors and the \swift\ spacecraft ACS.

\section{Latency Budget}
\label{sec:latency}
Thus far we have focused on irreducible, fixed, latencies originating from the evolution of the GW signal given the current IGWN sensitivity, the size of the BAT FOV, and the attitude control system of \swift. However, there are significant non-intrinsic additional latency overheads associated with both.
\subsection{IGWN}
We have so far assumed instantaneous access to the information contained in the GW data at a given time/truncation frequency, ignoring the associated latencies of recording the data at the interferometers, calibration, data transfer to cluster, search pipeline filtering, parameter estimation (including sky location), and alert distribution. \citet{O4latency} characterize the latency of the GW alert products via a Mock Data Challenge, but do not include latency originating before the data is available for searches. However, they do note that the total measured latency from real data, characterized as GW event time (merger) to alert \textit{production}, has a median of 29.7 s in O4a. This is comprised of $\sim5-10$ s latency in data acquisition, calibration and transfer to computing cluster, $\sim5-10$ s for search pipelines to find an event and trigger, $\sim1$ s for sky map creation, and another $\sim7-10$ s for parameter estimation, annotation, and alert orchestration.

\begin{figure}
    \centering
    \includegraphics[width=0.5\linewidth]{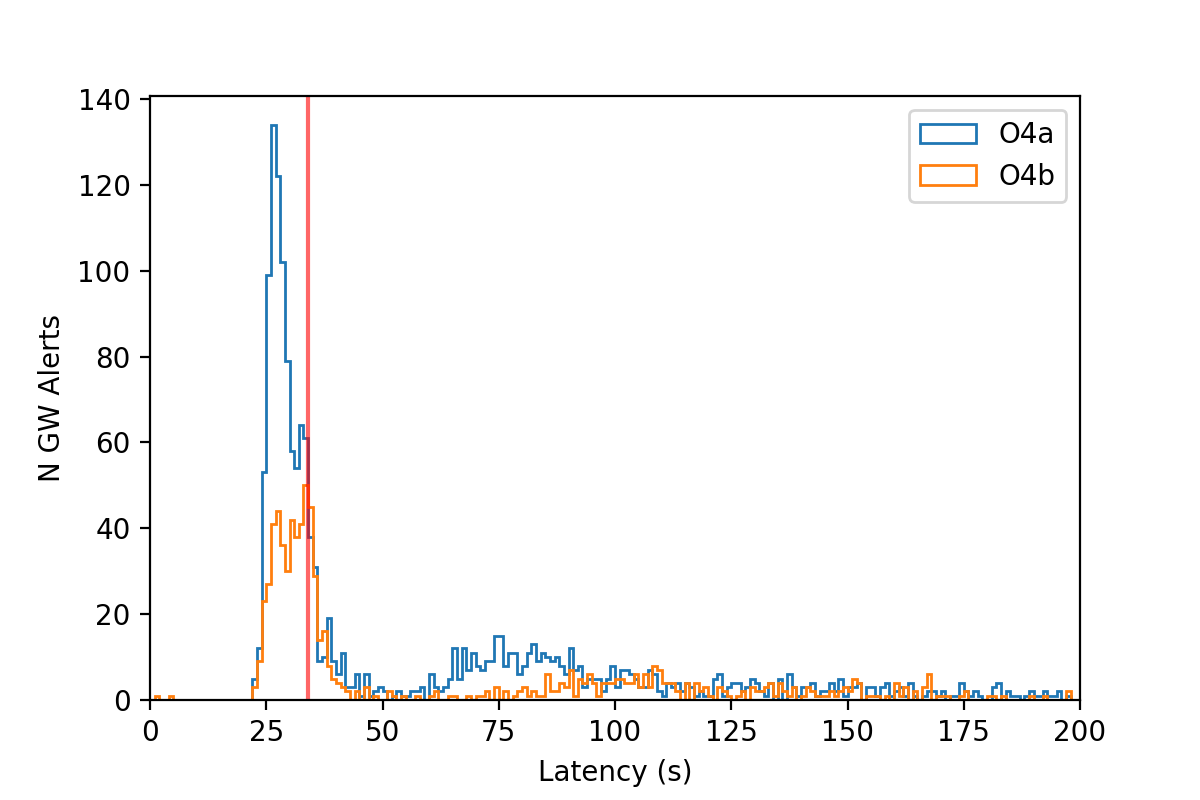}
    \caption[The measured latency of GW alerts during Observing Run 4.]{The latency of GW alerts during O4, measured as the difference between alert receipt and merger time. The red line denotes the median latency across all alerts in O4. Note that these latencies must be added on top of early warning negative latencies, leading most `early warning' alerts to be distributed after merger.}
    \label{fig:GWlatencies}
\end{figure}

We repeat this end-to-end latency measurement for all of O4 (to date), measuring the latency directly as the difference between the merger time and \textit{received} alert time. We purposefully screen out the early warning alerts (all thus far retracted) so as to reliably measure the full-stack latency from GW signal in the detector to alert. This latency can then be safely added on top of any CBC analysis, including early warning. Fig. \ref{fig:GWlatencies} shows the latency distribution for both O4a and O4b. We measure a median latency of $\sim34$ s, higher than presented in \citet{O4latency}. This latency overhead is comparable to the early warning achievable at current detector sensitivity. \textbf{We highlight that the current pipeline latency mostly obviates the effectiveness of the early warning alerts at providing actionable information before merger.}

\subsection{\swift}
On the followup side we have considered only the slew latencies of the \swift\ ACS, assuming instant slew start. We have so far ignored the ground processing of the alert, decision latency, and telecommand latency to the spacecraft. In \citet{guano} we presented a novel pipeline that reduced the on-demand command latency of \swift\ from 4 hours to 14 minutes, measured as the difference between alert receipt on the ground and command receipt on orbit.
The majority of this 14 minute latency floor stemmed from an immutable characteristic of on-demand scheduling of the Tracking Data and Relay Satellite System (TDRSS), which does not permit new contacts scheduled autonomously to begin less than 10 minutes from request time to ensure adequate time for service configuration changes between different Space Network (SN) users \citep{christdrss}.

However, once the Swift Mission Operations Center (MOC) and the satellite are connected, the command latency through TDRSS should be very low. We can break this into a few fundamental components, tracing the journey of a command bit from the \swift\ MOC to \swift\ on orbit:
\begin{enumerate}
    \item Terrestrial network delays, MOC to SN ground system.
    \item Propagation delay through SN ground terminal.
    \item Time of flight from SN ground terminal to TDRS satellite (space-to-ground link, SGL).
    \item Propagation delay through the TDRS satellite.
    \item Time of flight from TDRS satellite to \swift\ (space-to-space link, SSL).
    \item \swift\ receiver waiting for the entire command telemetry frame, at our data transfer rate.
\end{enumerate}

\noindent The sum of these latencies is conservatively less than 2 s! However, achieving this would require \swift\ to already be in a scheduled contact at the unknown future time of a neutron star merger.

Motivated by the unique scientific opportunity presented by early warning GW detection, and the high instantaneous availability fraction ($\geq85\%$) of TDRSS Multiple Access Forward service demonstrated by \citet{guano} and \citet{christdrss}, the Space Network has enabled the \swift\ mission to pre-schedule semi-continuous contact service when available. After several test periods, this service began ongoing operation in Spring 2023, now achieving an average availability of $\sim80\%$. We call this capability `continuous commanding.' Effectively, we are now always in touch with \swift.\footnote{It would have been much easier if our space telescope were in GEO...}

After the spacecraft has received the slew request, it determines if the requested slew is safe to begin immediately. This takes $<1$ second. If so, it notifies the instruments that a slew is about to begin, and will wait up to 2 s for UVOT to reply with an acknowledgement, before beginning to slew. As such, we should expect an absolute minimum of 3-5 s latency from command sent on the ground until slew start.

\subsection{\swift\ Demonstration}
In order to characterize the true performance of this new system, and exemplify broader application to time-domain science, we demonstrate \swift\ response to an external Fast Radio Burst (FRB) trigger received from CHIME. This FRB arrived on Earth at T0=2023-09-05T15:18:46.48 UTC (arrival time dedispersed to infinite frequency), the VOEvent reporting this detection was received at the \swift\ MOC 46 s later, at T$_{\mathrm{alert}}$=15:19:33. Here we provide a subsequent timeline for \swift\ repointing:

\scalebox{1}{
\begin{tabular}{r |@{\foo} l}
T$_{\mathrm{alert}}+31$ s & TOO command sent from MOC to \swift\\
T$_{\mathrm{alert}}+40$ s & Slew start, 81 degrees to go.\\
T$_{\mathrm{alert}}+88$ s & Target position enters BAT FOV. \\
T$_{\mathrm{alert}}+149$ s & XRT and UVOT begin recording photons from target position.\\

\end{tabular}
}

\textit{Note the 9 second delay between sending the command and \swift\ beginning to slew!} Compare this to the previous best performance of on-demand response, at 10s of minutes. In this case the 31 second delay between alert received and command sent originates from a sub-optimal ingestion and safety check procedure in the \swift\ ground system. Using \textbf{the new procedure described in previous sections reduces the latency from receipt to command being sent to 1 s}. Continuous commanding has since been integrated into general \swift\ operations, significantly increasing observatory versatility and efficiency. It has been used to send hundreds of TOO commands to date. \textbf{The median delay between command sent and slew start is presently $10$ s, and the minimum achieved is $6$ s.} We continue to investigate ways to further decrease this latency. 

\section{Conclusions}
\label{sec:conclusions}
Exceptional scientific yield is possible with early-time ($<1$ minute) access to the post-merger environment. Prompt GRB detection and arcminute localization is presently the most promising avenue to achieve this, in the absence of decihertz GW detectors. \swift's superior sensitivity and precision positions are the best suited among current instruments, but are balanced with a FOV that covers only 1/6 of the sky, significantly suppressing detection rates given the low rate of BNS mergers in the local Universe. Undaunted, we investigate the question of whether \swift\ can point at the location of a BNS before merger by utilizing early warning GW detections.

We apply our \swift\ ACS model and trajectory optimizer to a large set of realistic simulations of early warning gravitational wave detections and sky maps. \swift\ is indeed capable of exploiting early warning gravitational wave alerts, slewing rapidly across the sky to put the GW location in the BAT FOV by merger time. With 60s of early warning this will more than double the rate of BNS mergers with prompt arcminute positions. While the localization precision increases with later warning time, for \swift\, and with the current IGWN sensitivity, responding to the earliest alert is the optimal strategy. Given the finite \swift\ slew speed, higher latency alerts result in a significant reduction in the expected yield. 

The IGWN is now sufficiently sensitive that the practical latencies between GW arrival in the interferometers and information reaching the \swift\ spacecraft now limit capabilities. This is a good thing, because in most cases cyber-infrastructure is easier to improve than GW interferometers or spacecraft already in orbit. However, we need to do it! 

We demonstrate a new \swift\ operational capability whereby we can begin slewing the spacecraft within 10 s of an external Target of Opportunity being received on the ground. With this innovation, the prompt arcminute detection rate is now fundamentally limited by the \swift\ slew speed (which cannot be improved) and the IGWN alert latency and software infrastructure (which can).  However, the IGWN alert latency is presently $\sim30$ s, \textbf{mostly obviating the practical utility of early warning alerts.} 

Several authors have outlined prospects for improved early warning capability with upgraded sensitivity of future detectors \citep{2022ApJ...935..139M, 2020ApJ...902L..29N}, as well as novel statistical approaches \citep{2023ApJ...959...76C,aframe}. We point out that in the current IGWN  implementation, the search pipelines themselves only contribute $\sim1/3$ of the total internal latency of the system, and co-equal focus must be placed on latency of data transfer, calibration (or not), and alert orchestration. We recommend significant focus on reduction of IGWN internal latencies to allow for this extremely compelling scientific opportunity.

In addition to GW early warning response with BAT, `continuous commanding' opens new opportunities for science with \swift. Upcoming surveys like CHORD \citep{chord}, the Argus Array \citep{argus1,argusdata}, LAST \citep{last}, and ULTRASAT \citep{ultrasat} will survey the sky at very high cadence, and distribute transient alerts at high rates and low latency. We anticipate significant demand from transients discovered by these surveys. This system is already being utilized by a community program for characterization of infant supernovae discovered by the Zwicky Transient Facility \citep{ztf} and BTSBot \citep{btsbot}. The \swift\ Team invites the community to consider and propose novel scientific applications of ultra-low latency UV, X-ray, and gamma-ray observations.

\begin{acknowledgments}
We thank the \swift\ Flight Operations Team for their invaluable assistance bringing continuous commanding online, and for consistent safe stewardship of \swift. We further thank Eric Siskind, Jamie Rollins, Alex Nitz, Chad Hanna, Reed Essick, Chris Matzner, Maria Drout, and many others for enlightening discussions.
\end{acknowledgments}

\vspace{5mm}
\facilities{Swift, LIGO, Virgo}


\software{  ligo.skymap, swift-tools, optuna \citep{2019arXiv190710902A}, astropy \citep{astropy:2013,astropy:2018}}



\bibliography{main}{}
\bibliographystyle{aasjournal}



\end{document}